\documentclass[prd, aps, reprint, amsfonts, amssymb, amsmath, preprintnumbers, showpacs, nofootinbib, superscriptaddress]{revtex4-1}
\usepackage{graphicx, multirow,soul}
\usepackage[citecolor=blue]{hyperref}
\usepackage[all]{hypcap}
\usepackage{url}
\usepackage{color}
\usepackage{subfigure}
\usepackage{slashed}
\usepackage{float}
\usepackage{bbm}
\usepackage{bbold}
\usepackage{lipsum}
\newcommand{\equaref}[1]{Eq.~(\ref{#1})}
\newcommand{\equasref}[2]{Eqs.~(\ref{#1})~and~(\ref{#2})}

\newcommand{\figref}[1]{Fig.~\ref{#1}}

\newcommand{\secref}[1]{Section~\ref{#1}}

\newcommand{\bq}{\begin{eqnarray}}
\newcommand{\nq}{\end{eqnarray}}
\newcommand{\no}{\nonumber}
\newcommand{\tr}{\text{tr}}

\newcommand{\elb}{\bar{\ell}}

\begin{document}

\title{Leptogenesis via Varying Weinberg Operator: a Semi-Classical Approach}

\author{Silvia Pascoli}\email{silvia.pascoli@durham.ac.uk}
\affiliation{Institute for Particle Physics Phenomenology, Department of
Physics, Durham University, South Road, Durham DH1 3LE, United Kingdom.}
\author{Jessica  Turner}\email{jturner@fnal.gov}
\affiliation{Theoretical Physics Department, Fermi National Accelerator Laboratory, P.O. Box 500, Batavia, IL 60510, USA.}
\author{Ye-Ling Zhou}\email{ye-ling.zhou@durham.ac.uk}
\affiliation{Institute for Particle Physics Phenomenology, Department of
Physics, Durham University, South Road, Durham DH1 3LE, United Kingdom.}

\date{\today}

\begin{abstract}
In this Letter, we introduce leptogenesis via a varying Weinberg operator from a semi-classical perspective. This mechanism is motivated by the breaking of an underlying symmetry which triggers a phase transition that causes the coupling of the Weinberg operator to become dynamical. Consequently, a lepton anti-lepton asymmetry arises from the interference of the Weinberg operator at two different spacetime points. Using this semi-classical approach, we treat the Higgs as a background field and show a reflection asymmetry between the leptons and anti-leptons is generated in the vicinity of the bubble wall. We solve the equations of motion of the lepton and anti-lepton quasiparticles to obtain the final lepton asymmetry. 
\end{abstract}
\preprint{IPPP/18/66, FERMILAB-PUB-18-330-T} 
 \pacs{98.80.cq}
\maketitle

\section{Introduction}
The origin of tiny neutrino masses and the asymmetry between baryons and anti-baryons in the Universe are two  fundamental and open questions in particle physics. An important theoretical development linking both is baryogenesis via leptogenesis \cite{Fukugita:1986hr}, which applies the new physics motivated by tiny neutrino masses to generate an asymmetry between leptons and anti-leptons. This lepton asymmetry  is later converted into the baryon asymmetry via weak sphaleron processes. 

Recently, we proposed a new mechanism to generate the lepton asymmetry via the Weinberg operator \cite{Pascoli:2016gkf} (see also  \cite{Pascoli:2016tiv, Pascoli:2017glr}). This operator is given by  
\bq
\mathcal{L}_\text{W}=-\frac{\lambda_{\alpha\beta}}{\Lambda} \ell_{\alpha L}^{i} \varepsilon^{ij} H^j C \ell_{\beta L} ^k \varepsilon^{kl} H^l +\text{h.c}, \label{eq:Weinberg}
\nq
where $\ell_L = (\nu_L, l_L)^T$ in the $SU(2)_L$ gauge space, $\lambda_{\alpha\beta}=\lambda_{\beta\alpha}$ are effective Yukawa couplings with flavour indices $\alpha,\beta=e,\mu,\tau$ and $C$ is the charge conjugation matrix. We demonstrated that the dimension five Weinberg operator can play a  crucial role in leptogenesis without the need to specify the completion of this operator. 
It provides two ingredients for the leptogenesis recipe: 
\begin{itemize}
\item The Weinberg operator violates lepton number by two units and triggers lepton-number-violating (LNV) processes, including
\bq
& H^* H^* \leftrightarrow \ell \ell\,,\quad
\overline{\ell} H^* \leftrightarrow \ell H\,,\quad
\overline{\ell} H^* H^* \leftrightarrow \ell \,,\no\\
&\overline{\ell} \leftrightarrow \ell H H\,,\quad 
H^* \leftrightarrow \ell \ell H\,,\quad
0 \leftrightarrow \ell \ell H H,
\nq
and their CP conjugate processes, where $\ell$ and $H$ are the left-handed leptonic and  Higgs doublet of the Standard Model, respectively. The CP violating phase transition occurs at much higher temperatures than the electroweak (EW) scale,
and therefore the Higgs has not acquired a non-zero vacuum expectation value (VEV) and it is almost in thermal equilibrium. 
\item
After electroweak symmetry breaking (EWSB), the Higgs acquires a VEV $\langle H \rangle = (0, v_H/\sqrt{2})^T$ and the
 neutrino mass matrix is given by
\bq 
(m_{\nu})_{\alpha\beta} = \frac{\lambda_{\alpha\beta}}{\Lambda} v_H^2 \,. 
\nq 
 This operator violates  lepton number and generates Majorana masses for neutrinos.  As the primary motivation for the Weinberg operator is the generation of tiny neutrino masses, all processes triggered by this operator are very weak \cite{Weinberg:1979sa}. The rate of these LNV processes is approximately
\bq
\Gamma_\text{W} \sim \frac{3}{4\pi^3}\frac{m_\nu^2}{v_H^4} T^3 \,,
\nq
where $v_H = 246$ GeV is the Higgs VEV and $m_{\nu} \lesssim 0.1$ eV is the neutrino mass. For temperatures $T<10^{13}$ GeV, as $\Gamma_\text{W}$ is smaller than the Hubble expansion rate, $H \sim \mathcal{O}(10) \frac{T^2}{m_\text{Pl}}$, the LNV processes generated by the Weinberg operator are out of thermal equilibrium. Moreover,  because of the smallness of the LNV rates, the washout mediated by the dimension-five operator is highly suppressed and can be safely ignored. 

\end{itemize}
In our mechanism, CP violation is provided by a CP-violating phase transition (CPPT) in the very early Universe.
This phase transition causes the coefficient of the Weinberg operator to be dynamically realised and contain irremovable complex phases.
Such a  phase transition is strongly motivated by a variety  of new symmetries such  as  $B-L$ and flavour symmetries.  In order to generate sufficient baryon asymmetry, we found the temperature of the 
 phase transition  to be approximately $10^{11}$ GeV. 
We discussed this mechanism in  \cite{Pascoli:2016gkf} and calculated the lepton asymmetry using non-equilibrium field theory methods. 
 Moreover, in our twin paper \cite{Pascoli:2018xx} we provide some additional discussion on the influence of the phase transition dynamics and how the particle thermal properties contribute to the mechanism.
 
  In this Letter, we present a simplified and intuitive description of this mechanism based on a semi-classical approximation. In order to do so, we follow the 
 method introduced in \cite{Huet:1994jb} where they calculated the  transition between left-and right-handed fermions via a varying mass during the electroweak phase transition (EWPT) \footnote{This work, along with several others \cite{Gavela:1993ts,Gavela:1994dt}, demonstrated that the amount of CP violation within the Standard Model (SM) is not sufficient to 
produce the observed baryon asymmetry of the Universe (BAU).}. The techniques applied in \cite{Huet:1994jb} are particularly amenable as the baryon asymmetry is calculated from  solving the equations of motions of the Green's functions of the left- and right-handed quasiparticles where the asymmetry itself manifests from the CP violating reflections of particles off the bubble wall.  The calculation is rather transparent and some of the simplifying assumptions they made, such as a thin and fast moving bubble wall, parallel our own. 

We emphasise that the CPPT mechanism works only if the UV-completion scale, $\Lambda$, is higher than the temperature of the phase transition $T$. If $\Lambda\lesssim T$, new lepton-number-violating particles, for example,  right-handed neutrinos needed for the type-I seesaw mechanism, may be produced in the thermal bath during the phase transition. Subsequently,  the phase transition may influence the leptogenesis via the decays of these particles as is studied in \cite{Pilaftsis:2008qt}.

We organise the remainder of this Letter as follows: we first review the mechanism in \secref{sec:mechanism};  we then  state the main assumptions of the semi-classical description in \secref{sec:approximation}.  Finally, we  present the calculation of lepton asymmetry in  \secref{sec:CPV} and make concluding remarks in \secref{sec:conclusion}.
\begin{figure}[t]\label{fig:bubble}
\centering
\includegraphics[width=0.48\textwidth]{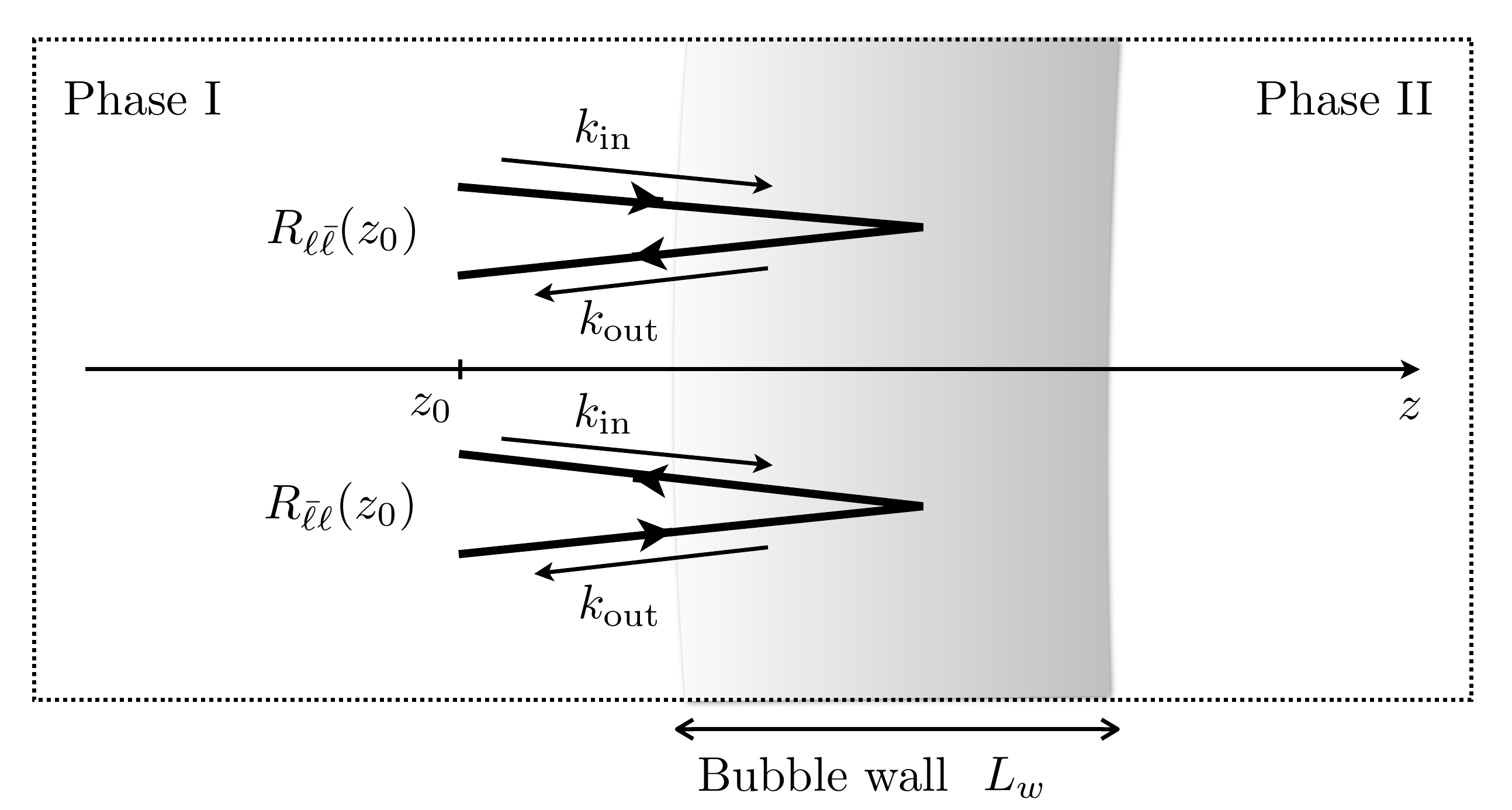}
\caption{\label{fig:bubble} Lepton and antilepton reflection on the bubble wall during phase transition from Phase I ($\langle\phi\rangle = 0$) to Phase II ($\langle\phi\rangle = v_{\phi}$) in the rest bubble wall frame. We set the bubble wall perpendicular to the $z$ direction. $R_{\ell \elb}(z_0)$ and $R_{\elb \ell}(z_0)$ represents the $z$-dependent transition amplitudes for lepton to anti-lepton and anti-lepton to lepton, respectively, at $z=z_0$.} 
\end{figure} 
\section{The CPPT Mechanism \label{sec:mechanism}}

The Weinberg operator of Eq.~\eqref{eq:Weinberg}  is the simplest higher-dimensional operator needed to explain tiny neutrino masses. As discussed in Refs.~\cite{Pascoli:2016gkf,Pascoli:2018xx}, in many models, the coupling of the Weinberg operator can be functionally dependent upon a SM-singlet scalar, $\phi$, such that $\lambda_{\alpha\beta}=\lambda^{0}_{\alpha\beta}+\lambda^{1}_{\alpha\beta}\langle\phi\rangle/v_{\phi}$.  Associated to $\phi$ is a finite temperature scalar potential, which is symmetric under a $U(1)_{B-L}$ or  flavour symmetry at sufficiently high temperatures.  As the temperature of the Universe lowers, the minima at the origin of this potential becomes metastable and a phase transition occurs. As a result, the minima changes from the vacua at the origin to a deeper, true vacua which is stable and non-zero, $\langle \phi\rangle$, and activates the CP violating coupling coefficient, $\lambda_{\alpha\beta}$.  The ensemble expectation value (EEV) of $\phi$ spontaneously breaks the high-scale symmetry and, if it is a flavour symmetry,  results in the  observed pattern of  leptonic masses and mixing. 

Assuming a first order phase transition, bubbles of the leptonically CP-violating broken phase  nucleate.
We denote the bubble wall width and bubble wall velocity as as $L_w$ and $v_w$, respectively.
In the following calculation, we work within the bubble wall rest frame where the bubbles wall is stationary
and the thermal plasma moves against the wall with velocity $-v_w$.
  Inside the bubble wall, the EEV is spacetime-dependent and therefore the coupling of the Weinberg operator, $\lambda_{\alpha\beta}$, must also vary with spacetime. This has the effect that the interference of the Weinberg operator at different times produces a lepton asymmetry. 

Before the CPPT is triggered, there are equal amounts of leptons and anti-leptons in the thermal plasma and they are thermally distributed. 
Once the CPPT begins, 
a bubble nucleates with the bubble wall separating the symmetric and broken phase
which are denoted in  \figref{fig:bubble}  as Phase I and II respectively. 
The majority of the leptons, anti-leptons and Higgses pass through the bubble wall; however, there will be some 
of these particles species which reflect off the wall. 
As the bubble wall causes the coupling of the Weinberg operator to be CP violating, the transition from leptons to anti-leptons and that from anti-leptons to leptons will be different in the presence of the bubble wall. Therefore different amounts of anti-lepton and leptons will be produced after these scatterings. As discussed, the interactions mediated by the Weinberg operator are out of thermal equilibrium and therefore LNV processes do occur but are rather rare.

We note that the coefficient of the Weinberg operator varies only along the $z$ direction in the wall as shown in  \figref{fig:bubble}.  Since the PT scale is much higher than the electroweak scale, all particles are massless, and the lepton $\ell$ and anti-lepton $\elb$ have helicity $-1$ and $+1$, respectively.
To further elaborate, we consider a group of leptons, $\ell$, propagating to the wall from the left hand side (Phase I). While most of the particles move freely through the wall to the right hand side (Phase II) without reflecting off the wall, a small proportion of the leptons will hit  the wall and subsequently convert to anti-leptons via the Weinberg operator. Since leptons and anti-leptons have opposite helicities, $\elb$ should move backwards to the Phase I zone. This process leads to the non-conservation of the momentum in the $z$ direction.

We denote the amplitude for transition from lepton to anti-lepton at $z=z_0$ as $R_{\ell\elb}(z_0)$. Likewise, for anti-lepton to lepton at $z=z_0$ we denote this amplitude as $R_{\elb\ell}(z_0)$. These transitions originate from the varying Weinberg operator and the CP asymmetry between these two processes is given by
\bq
\Delta_{CP}(z_0) \equiv |R_{\elb \ell}(z_0)|^2 - |R_{\ell \elb}(z_0)|^2\,. \label{eq:CP_def}
\nq
The interference of Weinberg operator at different $z$ can lead to {non-zero CP violating effects in the thermal plasma. 
In this case, given an equivalent amount of initial leptons and anti-leptons propagating from the right hand side; a different amount of anti-leptons and leptons can be generated via the reflection. This asymmetry finally diffuses to Phase II and will be preserved. The number density asymmetry of lepton and antilepton is given by
\bq
\Delta n_{\ell} = \int\frac{d^3k}{(2\pi)^3} [ f_\ell (k) - f_{\elb} (k) ]= \int\frac{d^3k}{2\pi} f_{\rm th} (k) \Delta_{CP}(z_0) \,, \no\\
\nq 
where  $f_{\rm th} (k) = \big[\exp(\beta\frac{\omega_k-v_w k_z}{\sqrt{1-v_w^2}}) + 1\big]^{-1}$ is the Fermi-Dirac thermal distribution boosted to the wall frame. 

In the following, we will carry out the semi-classical approximation to relate $\Delta_{CP}(z_0)$ with the varying Weinberg operator. 

The CPPT mechanism shares a common feature with EWBG; namely that a phase transition is 
necessary to drive the generation of a baryon asymmetry. However, the two mechanism differ markedly and it is 
worthwhile to remark on the features which distinguish them. First, in EWBG, the baryon number violation is provided by sphaleron transitions in the symmetric phase. Both the out-of-equilibrium condition and C/CP violations are induced by EW phase transition. Therefore, in EWBG, the phase transition is key to the generation of the non-equilibrium evolution. In order to achieve this, rapidly expanding bubble walls  are required such that the back-reactions are not efficient to wash out the generated baryon asymmetry. 
In the CPPT mechanism, the $B-L$ number violation and departure from thermodynamic equilibrium are directly provided by the very weakly coupled Weinberg operator. 
The PT is only necessary to provide a source of  C/CP violation and is not needed for 
the efficiency of reactions in the system. Consequently,  successful leptogenesis in this setup does not  necessarily require a  first-order PT and it is possible a CP-violating second-order PT would also  generate a lepton asymmetry. The purpose of assuming the first-order phase transition in this work is to simplify the discussion.

\section{The semi-classical approximation \label{sec:approximation}}
In this section, we introduce the semi-classical approximations we use for the lepton asymmetry calculation. Firstly,
we introduce the equations of motion (EOM) for the leptonic doublets and the effective mass-like matrix which parametrises the lepton anti-lepton transitions.  Secondly,  we review our treatment of the Higgs as a background field.

\subsection{Equation of Motion for Leptonic Quasiparticles}\label{sec:EOM}

We begin from the well-known equation of motion for Majorana neutrinos at low energy. It is expressed as
\bq 
\begin{pmatrix} i \sigma^\mu \partial_\mu & m_{\nu} \\ m_{\nu}^\dag & i \bar{\sigma}^\mu \partial_\mu \end{pmatrix}
\begin{pmatrix} \chi_{\nu} \\ \chi_{\bar{\nu}}
\end{pmatrix}
 = 0 \,,
\nq
where $\nu^c_L \equiv C \overline{\nu_L}^T = (\nu^c)_R$. The Majorana mass matrix, $m_\nu$, results in the neutrino anti-neutrino transitions and oscillations (see, e.g., \cite{Xing:2013woa}).

In the early Universe, when the Higgs is in its symmetric phase, the Higgs field may fluctuate. 
 Such fluctuations can be enhanced by temperature and influence the behaviour of neutrinos, as well as the charged leptons. For this reason, we 
 treat the Higgs as a background field.  Taking account the $SU(2)_L$ symmetry  the effective EOM for the leptonic doublet quasiparticles is directly obtained from \equaref{eq:Weinberg} as
\bq 
\begin{pmatrix} 
 i \sigma^\mu \partial_\mu & M_{\ell}(x) \\ M_{\ell}^\dag(x) & i \bar{\sigma}^\mu \partial_\mu
\end{pmatrix}
 \begin{pmatrix} 
  \chi_{\ell}(x) \\ \chi_{\bar{\ell}}(x) 
 \end{pmatrix} = 0 \,. 
\nq
In the $SU(2)_L$ gauge space the wave functions and mass-like matrix are given by
\bq 
\hspace{-8mm}
\chi_{\ell}(x) &=& 
\begin{pmatrix} 
 \chi_{\nu}(x) \\ \chi_{l}(x)
 \end{pmatrix} \,,\quad
\chi_{\bar{\ell}}(x) = 
\begin{pmatrix} 
-\chi_{\bar{\nu}}(x) \\ \,\,\chi_{\bar{l}}(x)
 \end{pmatrix}  \,, \\
\hspace{-8mm}
M_{\ell}^\dag(x) &=& \frac{\lambda(x)}{\Lambda}
\begin{pmatrix} 
2 \left[H^0(x)\right]^2 & \!-2 H^0(x) H^+(x)\! \\ \! -2 H^0(x) H^+(x)\! & 2 \left[H^+(x)\right]^2
 \end{pmatrix} \,,
\nq
where we have made the $x^\mu$-dependence explicit  to emphasise the spacetime-dependence of $M_\ell$. Note that the effective Majorana mass-like matrix,  $M_{\ell}(x)$, originates  from the Weinberg operator and leads the transition between lepton and anti-lepton which will be of importance for the lepton asymmetry generation.

\subsection{Higgs as a Background Field}\label{sec:HiggsBG}

As the Majorana mass-like matrix, $M_\ell(x)$, derives from the Higgs field, the thermal properties of this
scalar field will be of fundamental importance to the semi-classical treatment we detail in this paper.
Above the EWSB scale, the mean value of the Higgs field may be zero at finite temperatures, $\langle H \rangle =0$.
However, the mean value of $\langle H^\dag H \rangle$ is non-zero and such fluctuations correspond to particle excitations and annihilations in the thermal plasma. 

As a complex field, the mean value is given by
\bq\label{eq:meanvalue2}
\langle H^{0*} H^0 \rangle& =& \langle H^{+*} H^+ \rangle = \frac{1}{2} \langle H^\dag H \rangle\nonumber
\\ 
& =&2 \int \frac{d^3 k}{(2 \pi)^3} \frac{1}{2 \omega} \frac{1}{e^{\beta \omega}-1}  = \frac{T^2}{12},
\nq
where we have ignored the effective thermal masses and chemical potential of the Higgs. It is worth noting that the mean values of $\langle (H^0)^2 \rangle$, $\langle (H^+)^2 \rangle$ and $\langle H^0 H^+ \rangle$ should be zero. As we shall see later, this property will be important in the enhancement of the lepton asymmetry production at high temperatures.

Another interesting property is that the mean value $\langle (H^\dag(x) H(x))^2 \rangle$ is correlated with $\langle H^\dag(x) H(x) \rangle$ 
\footnote{It is proved in the following. For a real scalar $\varphi_i$, $\langle \varphi^2_i \rangle = T^2/12$, $\langle \varphi^{2n}_i \rangle = (2n-1)!! \langle \varphi^2_i \rangle^n$ \cite{Kapusta:1981aa}. For a complex scalar $\Phi = (\varphi_1 + i \varphi_2)/\sqrt{2}$,  $\langle \Phi^* \Phi \rangle = \frac{1}{2} \langle \varphi_1^2+\varphi_2^2 \rangle = T^2/12$, $\langle (\Phi^* \Phi)^2 \rangle = \frac{1}{4} \langle (\varphi_1^2+\phi_2^2)^2 \rangle = \frac{1}{4} \langle \varphi_1^4+\varphi_2^4 + 2\varphi_1^2 \phi_2^2 \rangle =  \frac{1}{4}(3\langle \varphi_1^2 \rangle^2 + 3\langle \varphi_2^2 \rangle^2 + 2\langle \varphi_1^2 \rangle \langle \varphi_1^2 \rangle) = 2 \langle \Phi^* \Phi \rangle^2$.} 
by 
\bq\label{eq:meanvalue4}
\langle (H^{0*}(x) H^0(x))^2 \rangle &= &\langle (H^{+*}(x)H^+(x))^2 \rangle \nonumber\\
&=&\frac{1}{3} \langle (H^\dag(x) H(x))^2 \rangle = \frac{T^4}{72}.
\nq

The expectation values for $H$ and $H^\dag$ at different spacetimes give the Wightman propagators, e.g.,
\bq
\langle H^{0*}(x_2) H^0(x_1) \rangle = S^{<}_{H^0} (x_1,x_2)\,, \no \\
\langle H^0(x_1) H^{0*}(x_2) \rangle = S^{>}_{H^0} (x_1,x_2)\,.
\label{eq:Wightman}
\nq
For the detailed discussion on correlations between lepton asymmetry and Wightman propagators, please see Ref. \cite{Pascoli:2018xx}. In this paper, we will ignore the spacetime difference between $H$ and $H^\dag$. This treatment simplifies the discussion and is sufficiently good to derive the CP asymmetry 
qualitatively.

\section{Lepton asymmetry in the semi-classical approximation \label{sec:CPV}}

The concept of quasiparticles has been 
known for many decades \cite{Klimov:1982bv,Weldon:1982aq} and manifests as 
 particle properties become modified in medium; for example 
particles may acquire a different mass from that in vacua as a result of their interactions in plasma. In general, such properties
can be described by collective excitations or a quasiparticle description. These quasiparticles are characterised by their 
dispersion relation which gives their energy ($\omega$) as a function of their momentum ($\textbf{k}$). Moreover, a stable particle in vacuum may have a finite lifetime in medium and this corresponds to the quasiparticle  \emph{damping rate}, $\gamma$.  The damping characterises the degree of decoherence of particles and therefore  gives a measure of the spread in the particle energy due to their interactions in medium. We define the  decoherence length, $L$, similarly to \cite{Huet:1994jb}
\begin{equation}
L=\frac{v_g}{2\gamma}=\frac{1}{6\gamma},
\end{equation}
where $v_{g}$ is the group velocity of the quasiparticle. As the quasiparticles of interest in our mechanism are leptons, the  decoherence results mainly  from the electroweak gauge interaction.
In this case, as the quasiparticles have homogeneous distributions parallel to the wall, it is reasonable to restrict our attention to quasiparticles with momenta perpendicular to the bubble wall \cite{Huet:1994jb}. 
We move to the rest wall frame and expand $\ell_L$ and $\overline{\ell_L}$ by positive and negative frequencies in the spinor space, respectively. As left-handed particles, they can be parametrised as 
\bq
\hspace{-5mm}
\ell_L &=& \begin{pmatrix} \exp[-i (\omega t - k_{\text{in}} z)]\chi_{1\ell}(z) \\ 
\exp[-i (\omega t + k_{\text{out}} z)] \chi_{2\ell}(z) \\ 0 \\ 0 \end{pmatrix} \,, \no\\
\overline{\ell_L}^T &=& \begin{pmatrix} 0 \\ 0 \\ 
\exp[+i (\omega t - k_{\text{in}} z)] \chi_{1\elb}(z) \\ 
\exp[+i (\omega t + k_{\text{out}} z)] \chi_{2\elb}(z) \end{pmatrix} \,. 
\nq
Here, we have required $\chi_{1\ell}$ and $\chi_{1\bar{\ell}}$ to be incoming quasiparticles moving in the $+z$ direction, and $\chi_{2\ell}$ and $\chi_{2\bar{\ell}}$ to be outgoing quasiparticles moving in the $-z$ direction (\emph{i.e.} the quasiparticles in this upper component of the spinor are moving \emph{into} the bubble, and the lower component are reflected back to Phase I by the wall). $\chi_{1\ell}(z)$ and $\chi_{2\elb}(z)$ have spin $j_z=-\frac{1}{2}$, while $\chi_{1\elb}(z)$ and $\chi_{2\ell}(z)$ have spin $j_z=+\frac{1}{2}$. The coherence of these states may be included using the following  replacement
\bq
k_{\text{in}} &\to& K_{\text{in}} = k_{\text{in}} + \frac{i}{2 L} \,, \no\\
k_{\text{out}} &\to& K_{\text{out}} = k_{\text{out}} - \frac{i}{2 L},
\nq 
with $\gamma_w = \gamma \sqrt{1-v_w^2}$ being the boosted damping rate. 
As $M_\ell(z)$ does not change energy in the wall frame, we do not distinguish between the energy, $\omega$, of the leptons and anti-leptons.
The EOM is decomposed into two uncoupled equations, one for $j_z=-\frac{1}{2}$ and the other for $j_z=+\frac{1}{2}$ quasiparticles. They are expressed as
\bq
\hspace{-7mm}
\left[\left(-i\partial_z+ \omega \right)\mathbb{1}_2-
\begin{pmatrix} 
-K_{\text{in}} & M^\dag_{\ell}(z) \\ -M_{\ell}(z) & -K_{\text{out}}
\end{pmatrix} 
\right]
\begin{pmatrix} \chi_{1\ell}(z) \\ \chi_{2\bar{\ell}}(z)
\end{pmatrix}
&=& 0,
\label{eq:master1}
\\
\hspace{-7mm}
\left[\left(-i\partial_z- \omega \right)\mathbb{1}_2- 
\begin{pmatrix} 
 K_{\text{in}} & -M_{\ell}(z) \\ M^\dag_{\ell}(z) & K_{\text{out}}
 \end{pmatrix} 
 \right] 
 \begin{pmatrix} 
  \chi_{1\bar{\ell}}(z) \\ \chi_{2\ell}(z) 
   \end{pmatrix} 
   &=& 0,
\label{eq:master2}
\nq
respectively. The energy-dependent term does not contribute to the CP violation in the rest wall frame, and thus we will not include it in the following discussion. 

 The calculation of the lepton asymmetry generated from CPPT will follow from solving the 
EOMs for the leptonic doublet quasiparticles. 

Now we consider the amplitude of $\chi_{1\ell}$ transition to $\chi_{2\bar{\ell}}$ and use the techniques developed in \cite{Huet:1994jb} for electroweak baryogenesis (EWBG). The transition from left-handed fermion to right-handed fermions via a spacetime-varying mass is similar to our case of the transition from left-handed lepton to right-handed anti-lepton via the time-varying Weinberg operator. 

The first step is to consider the propagation of quasiparticles in Phase I where we restrict our
discussion to the $j_z=-1/2$ quasiparticles $\chi_{1\ell}$ and $\chi_{2\elb}$. The relevant Green functions are
\bq
\left(-i\partial_z + K^{\text{in}(\text{out})}\right) G_{\ell (\bar{\ell})} (z-z_0) = \mathbbm{1} \delta(z-z_0) \,.
\label{eq:Green1}
\nq
In order to require no sources of quasiparticles at spatial infinity, the boundary conditions 
\bq
G_{\ell} (-\infty) = G_{\bar{\ell}} (+\infty) = 0,
\label{eq:BC1}
\nq
are necessary. 

The solution of the Green functions with the relevant boundary conditions is given by
\bq
G_{\ell}  (z-z_0) &=& i \theta(z-z_0) e^{-iK_{\text{in}} (z-z_0)} \no\\
&=& i \theta(z-z_0) e^{-(z-z_0)/(2L)} e^{-ik_{\text{in}} (z-z_0)} \,,\no\\
G_{\elb} (z-z_0)& = & -i \theta(z_0-z) e^{-iK_{\text{out}}(z-z_0)}\no\\
& =& -i \theta(z_0-z) e^{-(z_0-z)/(2L)} e^{-ik_{\text{out}}(z-z_0)}\,.\no\\
\label{eq:Green1_sol}
\nq
The lepton quasiparticle will propagate from Phase I into Phase II. For this purpose, we may consider 
leptons with a $\delta$-function source at $z=z_0$ propagating into the bubble wall. The influence of the wall leads to an effective ``mass'' term, $M_\ell(z)$, as explained above and the evolution of quasiparticles is described by \equaref{eq:master1}. Taking advantage of  the Green function method, we obtain 
\bq
\chi_{1\ell}(z) &=& -i G_{\ell} (z-z_0) \chi_{1\ell}(z_0) \no \\
&&+ \int dz_1 G_{\ell} (z-z_1) M^\dag_{\ell} (z_1)\chi_{2\elb}(z_1) \,, \no \\
\chi_{2\elb}(z)&=& \int dz_1 G_{\elb} (z-z_1) [-M_{\ell} (z_1)]\chi_{1\ell}(z_1).
\nq
Since the Weinberg operator is relatively weakly coupled
to the thermal plasma, we ignore all corrections $\lesssim \mathcal{O}(M^2_\ell)$. Therefore, the amplitude for $\chi_{1\ell}(z_0) \to \chi_{2\elb}(z_0)$, $R_{\ell \elb}(z_0)$, corresponding to the reflection matrix $R_{LR}$ in \cite{Huet:1994jb}, is given by
\bq
R_{\ell \elb}(z_0) &=& i \int dz_1 G_{\elb} (z_0-z_1) M_{\ell} (z_1)G_{\ell} (z_1-z_0)\nonumber \\
&=& i \int_{0}^{+\infty} dz_1 e^{-z_1/L} e^{i k_{\text{out}} z_1} M_{\ell} (z_0+z_1) e^{-i k_{\text{in}} z_1}.\nonumber\\
\nq
We can calculate the amplitude for $\chi_{1\elb}(z_0) \to \chi_{2\ell}(z_0)$ by assuming a similar treatment
of a $\delta$-function source at $z_0$ and moving in the $+z$ direction.
The resultant reflection matrix $R_{\elb \ell}$, corresponding to the reflection matrix $\bar{R}_{LR}$ in \cite{Huet:1994jb}, is
\bq
R_{\elb \ell}(z_0)
&=& i \int_{0}^{+\infty} dz_1 e^{-z_1/L} e^{ i k_{\text{out}} z_1} M^\dag_{\ell} (z_0+z_1) e^{-i k_{\text{in}} z_1}\,.\nonumber
\\
\nq
Finally, we obtain the CP asymmetry of the amplitude (defined in \equaref{eq:CP_def}) as
\bq
&&\Delta_{CP}(z_0) 
= \int_{0}^{+\infty} dz_1 dz_2 e^{-(z_1+z_2)/L} e^{ i (k_{\text{out}}-k_{\text{in}}) (z_1-z_2)} \no\\
&& \times [M^\dag_{\ell} (z_0+z_1) M_{\ell} (z_0+z_2) - M_{\ell} (z_0+z_1) M^\dag_{\ell} (z_0+z_2)] \no\\
&&= 2 \int_{0}^{+\infty} dz_1 dz_2 e^{-(z_1+z_2)/L} \; \sin[(k_{\text{out}}-k_{\text{in}}) (z_1-z_2)] \no\\
&& \times \text{Im}[M_{\ell} (z_0+z_1) M_{\ell}^\dag (z_0+z_2)]
\label{eq:CP2}
\,.
\nq
This quantity is determined by 1) the momentum change  $k_{\text{out}}-k_{\text{in}}$ 
due to the pressure from the wall and 2) the  imaginary part of the interference of two varying Majorana mass-like matrices $\text{Im}[M_{\ell}^\dag (z_0+z_1) M_{\ell} (z_0+z_2)]$.
The criteria $\Delta_{CP} \neq 0$ at order $\mathcal{O}(M_\ell^2)$ can only be fulfilled if these two conditions are satisfied.
As discussed in \secref{sec:mechanism}, the $z$-dependent varying Weinberg operator can lead to momentum non-conservation in the $z$ direction. Ignoring the momentum exchange with the Higgs boson, this momentum non-conservation is explicitly written as
\bq
k_{\text{out}} &\neq& k_{\text{in}} \,. 
\nq
The momentum difference $k_{\text{out}} - k_{\text{in}}$ represents the impulse of the wall acting on the leptons and anti-leptons. A similar problem is encountered in EWPT studies and the on-shell condition is usually assumed, where the momentum difference is correlated with the mass varying along the $z$ direction. The on-shell condition is relaxed once transition radiations are included, and the latter is more important if bubble wall moves very fast \cite{Bodeker:2017cim}. In our case, applying the on-shell condition can only give us very small momentum change because  $M_\ell$ is very small. 
A large momentum change can be obtained through interactions of the scalar excitation with the leptons, anti-leptons and Higgses.
Such processes manifest as there is an energy gradient within the bubble wall and the scalar excitation can be produced off-shell and
interact with the leptons, anti-leptons  and Higgses thereby causing perturbations in their distribution functions from equilibrium. 
To simplify the problem, we make the reasonable assumption that the maximum value of the momentum transfer is of the order of the plasma temperature \cite{Pascoli:2018xx}.

We will discuss in detail the term $\text{Im}[M_{\ell}^\dag (z_0+z_1) M_{\ell} (z_0+z_2)]$ shown in the above expression. We can rewrite this term as $AB/\Lambda^2$, where $A$ and $B$ specify the flavour and gauge component contributions respectively.
 For CP violation between $\elb_\alpha \to \ell_\beta$ and its conjugate process to occur, we have  
\bq
A_{\alpha \beta} = \text{Im}\{ \lambda_{\alpha\beta}(z_0+z_1)\lambda^*_{\alpha\beta}(z_0+z_2)\} \,.
\nq
And the total contribution with all flavour summed together is given by 
\bq
A \equiv \sum_{\alpha\beta} A_{\alpha \beta} = \text{Im}\{ \text{tr}[\lambda^*(z_0+z_1)\lambda(z_0+z_2)] \} \,. 
\nq
For CP asymmetry between $\bar{\nu}-\nu$,  $\bar{l}-\nu$, $\bar{\nu}-l$, $\bar{l}-l$ transitions, $B$, is respectively given by
\bq
B_{\bar{\nu}\nu} &=& 4 (H^{0*}H^0)^2 \,, \no\\
B_{\bar{l}\nu} &=& 4 (H^{0*}H^0) (H^{+*}H^+) \,, \no\\
B_{\bar{\nu}l} &=& 4 (H^{0*}H^0)^2 \,, \no\\
B_{\bar{l}l} &=& 4 (H^{0*}H^0) (H^{+*}H^+) \,, 
\nq
Ignoring the energy-momentum exchange between leptons and Higgs, and taking mean values on the right-hand-side as proved in \equasref{eq:meanvalue2}{eq:meanvalue4}, we obtain 
\bq
B_{\bar{\nu}\nu} = B_{\bar{l}l}  = \frac{T^4}{18} \,, \quad
B_{\bar{l}\nu} = B_{\bar{\nu}l} = \frac{T^4}{36}  \,. 
\nq
The average among gauge components is given by
\bq
B &\equiv& \frac{1}{2} (B_{\bar{\nu}\nu}  + B_{\bar{l}\nu} + B_{\bar{\nu}l} + B_{\bar{l}l} ) = \frac{T^4}{12} \,.
\nq
Taking into account the results of $A$ and $B$ as given above, we obtain the integration of  \equaref{eq:CP2}. We find it is dependent upon three terms: the interference of coefficient  term $A$; the damping term $e^{-(z_1+z_2)/L}$ and the oscillation term $\sin[(k_{\text{out}} - k_{\text{in}})(z_1-z_2)]$. In general, the wall length and decoherence  length are inversely proportional to the temperature and the momentum transfer is proportional to the temperature. Therefore,  the CP asymmetry $\Delta_{CP}(z_0)$ in \equaref{eq:CP2} is proportional to $\text{Im}\{\tr[\lambda^0 \lambda^{1*}]\} T^2/\Lambda^2$, with the coefficient depending on the competition of the three terms, where $\text{Im}\{\tr[\lambda^0 \lambda^{1*}]\} / \Lambda^2 = \text{Im}\{\tr[m_\nu^0 m_\nu^{*}]\} /v_H^4$. 
Therefore, the final baryon asymmetry is given by 
\bq
\eta_B \sim \frac{\Delta n_{\ell}}{n_\gamma} &\sim& \text{Im}\{ \tr[m_\nu^0 m_\nu^*] \} \frac{T^2}{v^4_H} \,,
\nq
which is qualitatively the same as our previous result \cite{Pascoli:2016gkf}.
Through this simplified treatment, we recover the combination $\text{Im}\{ \tr[m_\nu^0 m_\nu^*] \}$ and the temperature-dependent contribution $\propto T^2$ to the number density asymmetry between lepton number and anti-lepton number. 

\section{Conclusion \label{sec:conclusion}}
In this paper, we apply a semi-classical approximation to calculate the lepton asymmetry generated by the varying Weinberg operator. Firstly, we approximate the Higgs field as a background field. Following this treatment, we can effectively regard the Weinberg operator as an effective ``Majorana mass term'' for the leptonic doublet. Then, we write out the EOM for both lepton and anti-lepton quasiparticles, in which the ``Majorana mass term'' results in lepton anti-lepton transition. During the CP-violating phase transition, the ``Majorana mass term'' varies with spacetime, and the transition from lepton to anti-lepton and that from anti-lepton to lepton are not equal. This treatment is analogous to one approximation used in EWBG, where the varying fermion mass result in the asymmetric transition between left-handed and right-handed components. 

In this semi-classical approximation, we do not try to provide quantitatively precise results of the lepton asymmetry as the energy-momentum transfer with the Higgs has been ignored. However, this simplified treatment allows us to present the mechanism more intuitively. 
Moreover, one of the main results of this paper is that, in the single scalar case, the number density asymmetry between lepton and anti-lepton $\Delta n_{\ell} \propto \text{Im}\{ \tr[m_\nu^0 m_\nu^*] \} T^2/ v^4_H$ agrees with the result obtained using the non-equilibrium QFT approach.
\\

\acknowledgements
This manuscript of S.P. and Y.L.Z. was supported by European Research Council under ERC Grant NuMass (FP7-IDEAS-ERC ERC-CG 617143), H2020 funded ELUSIVES ITN (H2020-MSCA-ITN-2015, GA-2015-674896-ELUSIVES) and InvisiblePlus (H2020-MSCA-RISE-2015, GA-2015-690575-InvisiblesPlus), 
and that of J.T. was authorised by Fermi Research Alliance, LLC under Contract No. DE-AC02-07CH11359 with the U.S. Department of Energy, Office of Science, Office of High Energy Physics. 
Y.L.Z. would like to give particular thanks to Z. z. Xing and the Institute of High Energy Physics, Chinese Academy of Sciences where the early stage of this work was carried out. S.P. acknowledges partial support from the Wolfson Foundation and the Royal Society. 

\bibliographystyle{apsrev4-1}
\bibliography{ref}{}
\end{document}